\documentstyle[preprint,eqsecnum,aps]{revtex}

\def\go{\mathrel{\raise.3ex\hbox{$>$}\mkern-14mu
             \lower0.6ex\hbox{$\sim$}}}
\def\lo{\mathrel{\raise.3ex\hbox{$<$}\mkern-14mu
             \lower0.6ex\hbox{$\sim$}}}
\def\be{\begin{equation}}
\def\ee{\end{equation}}
\def\br{{\bf r}}
\def\bR {{\bf R}}

\def\bB {{\bf B}}
\def\bK {{\bf K}}
\def\bp {{\bf p}}
\def\bA {{\bf A}}
\def\bPi {{\bf \Pi}}
\def\cE {{\cal E}}
\def\cA {{\cal A}}

\begin{document}
\draft
\title{
Motion and Ionization Equilibrium of Hydrogen Atoms\\
in Superstrong Magnetic Field}

\author{Dong Lai}
\address{Theoretical Astrophysics, 130-33, 
California Institute of Technology\\
Pasadena, CA 91125\\
{\rm E-mail: dong@tapir.caltech.edu}}
\author{Edwin E.~Salpeter}
\address{
Center for Radiophysics and Space Research, 
Cornell University\\ 
Ithaca, NY 14853\\
\vskip 0.4in}

\date{To be published in {\it Physical Review A}, 1995} 

\maketitle 
\begin{abstract}

We study the effects of finite proton mass on the energy levels
of hydrogen atoms moving transverse to a superstrong magnetic field
$B$ with generalized pseudomomentum $K_\perp$. Field strengths of order
$B\sim 10^{12}$ Gauss are typically found on the surfaces of neutron
stars, but we also study the regime $B\go B_{crit}= 4.23\times
10^{13}$ Gauss, where the Landau excitation energy of the proton is large. 
We adopt two different approaches to the two-body problem in strong 
magnetic field, and obtain an approximate, but complete solution 
of the atomic energy as a function of $B$ and $K_\perp$.
We show that, for $B>>B_{crit}$, there is an orthogonal set 
of bound states, which do not have any Landau excitation
contribution in their energies. 
The states with very large $K_\perp$ have small binding energy
and small transverse velocity, but are nevertheless
distinct from the fully ionized states. The final results for 
the excitation energies are given in the form 
of analytical fitting formulae. 

The generalized Saha equation for 
the ionization-recombination equilibrium of hydrogen gas 
in the presence of a superstrong magnetic field is then derived. 
Although the maximum transverse velocity of a bound atom decreases
as $B$ increases, the statistical weight due to transverse motion 
is actually increased by the strong magnetic field. 
For astrophysically interesting case of relatively low density and
temperature, we obtain analytic approximations for 
the partition functions. The highly excited bound states have 
a smaller statistical weight than the fully ionized component.

\end{abstract}
\bigskip
\pacs{32.60.+i, 97.10.Ld, 31.20.Di, 97.60.Jd}

\bigskip

\section{\bf INTRODUCTION}
\label{sec:intro}

It is well known that a superstrong magnetic field of 
order $B\go 10^{12}$ Gauss, typically found on the 
surfaces of neutron stars, 
can dramatically change the structure
of neutral atoms and other bound states 
(see, e.g., Ref.\cite{Ruderman74} for an early review and
Ref.\cite{Ruder94} for a recent text on atoms in strong magnetic fields). 
The atomic unit $B_o$ for the magnetic field strength and a dimensionless
parameter $b$ are
\be
B_o={m_e^2e^3 c\over\hbar^3}=2.35 \times 10^9~{\rm G};~~~~
b\equiv {B\over B_o}.
\ee
When $b>>1$, the cyclotron energy of the electron
$\hbar\omega_e =\hbar (eB/m_ec)=11.58 B_{12}~{\rm keV}$,
where $B_{12}$ is the magnetic field strength in units of 
$10^{12}~{\rm G}$, is much larger than the typical Coulomb energy, 
thus the Coulomb forces act as a 
perturbation to the magnetic forces on the electrons, and 
at most temperatures
the electrons settle into the ground Landau level. 
Because of the extreme confinement of electrons in the
transverse direction, the Coulomb force becomes 
much more effective for binding electrons in the parallel direction. 
The atom has a cigar-like structure. Moreover, it is possible for 
these elongated atoms to form molecular chains by covalent bonding along 
the field direction.
 
Hydrogen atoms in strong magnetic field have been studied 
extensively\cite{Elliot60,Cohen70,Garstang77,Rosner84,Goldman91}.
We have also recently completed a study of the electronic structure 
of hydrogen molecules and chains 
in the strong field regime ($b>>1$)\cite{Lai92}. 
However, in many of these studies, 
the center-of-mass (CM) effects of the proton motion have been neglected. 
In the case of the hydrogen atom, although 
significant effort has been devoted to calculating 
the energy levels of an electron in the static Coulomb potential of a
{\it fixed\/} proton (infinite mass) 
to a high precision and for arbitrary magnetic field
strength\cite{Rosner84,Goldman91}, 
the two-body problem in strong magnetic field, including the effects 
of finite proton mass, has been
studied in detail only recently (e.g.,\cite{Vincke92,Potekhin94}).
In this paper we focus on the hydrogen atom, but discuss those 
aspects which are important for 
application to molecules in very strong fields (we shall study 
molecular excitation levels in a later paper\cite{Lai95}). 

A free electron confined to the
ground Landau level, the usual case for $b>>1$, 
does not move perpendicular to the magnetic field. Such motion
is necessarily accompanied by Landau excitations. 
When the electron (still in the Landau ground state) 
combines with a proton, the mobility of the neutral atom across the field
depends on the ratio of the atomic excitation energy and 
the Landau excitation energy $\hbar\omega_p=\hbar eB/(m_pc)$ 
for the proton. As the typical Coulomb excitation is 
$(\ln b)$ in atomic units, 
it is convenient to define a critical field strength $B_{crit}$ via
\be
b_{crit}\equiv {m_p\over m_e}\ln b_{crit}=1.80\times 10^4;
~~~~
B_{crit}=b_{crit}B_o=4.23\times 10^{13}~{\rm G}.
\ee
For $B<<B_o$ (terrestrial conditions) the Landau energies of both 
electron and proton are minor perturbations and one can construct wave 
packets that mimic the classical motion of a neutral atom across the field. 
In this case the internal structure of the atom or molecule is 
coupled to the center-of-mass motion only 
through a ``motional Stark effect'' (see Sec.~II). 
In the opposite limit of $B>>B_{crit}>>B_o$, the Landau energies are 
larger than the Coulomb excitation energy, and both electron and proton 
(in the atom) are in the Landau ground state at reasonable
temperatures. In this case, quantum mechanics cannot 
mimic classical motion. The canonical field strengths of radio pulsars, 
with $B$ slightly less than $B_{crit}$, present an intermediate case. 
However, at all field strengths one can introduce a pseudomomentum 
$\bK_\perp$ perpendicular to the field and, in principle, find 
the bound-state energy $\cE$ as a function of 
$\bK_\perp$. One question of interest is the 
range of $K_\perp$ for which $\cE$ increases linearly with $K_\perp^2$
(as it does for ordinary kinetic energy); but in any case $\cE(K_\perp)$
is needed to evaluate the Saha equation for the equilibrium between 
neutral and ionized hydrogen (Sec.~V). 
Previous treatments\cite{Khersonskii87,Miller92,Ventura92,Pavlov94}
of the ionization equilibrium in strong magnetic fields 
did not properly take account of the non-trivial effects of 
atomic motion. States with large $K_\perp$, where
velocity decreases with increasing $K_\perp$ (see
Ref.\cite{Potekhin94} and our Eq.~[3.29]) are of particular 
interest, especially for $B$ larger than $B_{crit}$. 

The separation of the center-of-mass (CM) motion of a bound state
in the presence of a magnetic field has been studied previously
based on the conserved pseudomomentum
(e.g.,\cite{Lamb52,Gorkov68,Avron78,Herold81,Johnson83,Schmel88}).
We briefly review and clarify this pseudomomentum scheme in Sec.~II. 
When $B<<B_{crit}$, perturbation calculations for 
hydrogen atom motion (e.g.,\cite{Herold81,Vincke88,Pavlov93})
are valid over a wide range of $K_\perp$ values and lead to 
interesting phenomena such as bent trajectories\cite{Pavlov93}.
Model atmospheres can be affected by details of the transverse 
motion (e.g.,\cite{Pavlov94b,Ventura94}).
Some accurate numerical calculations for general $K_\perp$ and for 
several values of $B$ (but all below $B_{crit}$) are now available
\cite{Vincke92,Potekhin94}, but we concentrate on 
the $B\go B_{crit}$ regime in Sec.~III-IV.
A different approach to the two-body problem (for positronium atom)
in the superstrong field regime $b>>1$ has been developed 
in\cite{Shabad86}.

Our purpose in this paper is not to obtain accurate energy spectra of 
a moving hydrogen atom in certain limited regimes, as have been done 
in most of the papers mentioned above; rather, 
we seek complete (though approximate) solutions of the two-body problem 
for a wide range of parameter space, 
including $B$ just below $B_{crit}$ (a common case for 
neutron stars), but especially for $B>B_{crit}$ (in case $B\go 10^{14}$ G 
exists in some neutron stars, as has been suggested 
recently\cite{Paczynski92,Duncan93,Chakrabarty94}). 
Our emphasis is on finding physically meaningful approximate 
fitting formulae for the atomic energy of the moving atom
over all relevant values of $K_\perp$ (Sec.~III), 
in order to determine the equilibrium 
between neutral and ionized hydrogen (Sec.~V).
However, in practice we shall be interested mainly in the regimes where 
the thermal energy $k_BT$ is much less than the ground-state binding 
energy of the atom, while the gas density is much smaller than the 
internal density of the atom, so that the neutral and ionized 
fractions are of the same order of magnitude. In such cases, 
we are most interested in two kinds of excited bound states: 
(i) Those with excitation energies up to a few atomic units
(comparable to $k_BT$ but a small fraction of the binding energy) and
(ii) states that are only barely bound
(e.g., those with extremely large $K_\perp$), for which one has to check 
whether phase-space factors make them unimportant relative to 
ionized hydrogen. 
For applications to molecules\cite{Lai92,Lai95} 
and multi-electron atoms\cite{Neuhauser87} with $B>>B_{crit}$,
a controversy arises regarding the ``coupling'' of the electron's
orbital quantum number with the Landau level of the proton
(or nuclei). This is discussed in Sec.~IV.

Our calculations in this paper are based on non-relativistic quantum 
mechanics. For $B\go B_{rel}
=(\hbar c/e^2)^2B_o=4.414\times 10^{13}$ G (note that $B_{rel}$ 
is close to $B_{crit}$ only by coincidence), i.e., 
$\hbar\omega_e\go m_ec^2$, the transverse motion 
of the electron becomes relativistic. However, the relativistic correction 
to the atomic binding energy is small as long as the electron remains
non-relativistic along field direction\cite{Angelie78}. 
Except as otherwise noted, 
we shall use atomic units (a.u.) throughout the paper, in which
mass and length are expressed in units
of the electron mass $m_e$ and the Bohr radius
$a_o=\hbar^2/(m_e e^2)=0.529\times 10^{-8}$ cm, energy in units of
$2~{\rm Rydberg}=e^2/a_o=2\times 13.6$ eV;
field strength is in units of $B_o$ (Eq.~[1.1]) and temperature
in units of $3.15\times 10^5$ K.

\section{\bf SEPARATION OF CENTER-OF-MASS MOTION: THE PSEUDOMOMENTUM 
APPROACH}
\label{sec:app}

To set the scene, we briefly review the pseudomomentum approach 
to the two-body problem of a hydrogen atom in a strong magnetic 
field\cite{Gorkov68,Avron78,Herold81}. However, the physical meaning
of the pseudomomentum of the atom needs some clarification. 

\subsection{\bf Pseudomomentum}

For a free particle of charge $e_i$ and mass $m_i$ in a constant
magnetic field (assumed to be aligned in the $z$-axis), 
there are three momentum-like vectors:
the {\it canonical momentum} ${\bf P}=-i{\bf\nabla}$;
the {\it mechanical momentum} $\bPi={\bf P}-e_i{\bf A}=m_i{\bf v},$
where ${\bf A}$ is the vector potential and ${\bf v}$ is the velocity;
and the {\it pseudomentum} (or the generalized momentum), as defined by
\be
{\bf K}=\bPi+e_i{\bf B}\times {\bf r}.
\ee
That $\bK$ is a constant of motion can be easily seen from the classical
equation of motion for the particle $d\bPi/dt=e_i(d\br/dt)\times\bB$.
The parallel component $K_z$
is simply the linear momentum, while the constancy of the perpendicular
component $\bK_{\perp}$ is the result of the fact
that the guiding center of the circular orbit of the particle
does not change with time. The position vector
${\bf R}_c$ of this guiding center is related to $\bK_{\perp}$ by
\be
\bR_c={\bK_{\perp}\times {\bf B}\over e_iB^2}.
\ee
Mathematically, the conservation of $\bK$ 
is the result of the invariance of the Hamiltonian under a
spatial translation plus a gauge transformation
\cite{Avron78}. 

The existence of the integration constant $\bK_{\perp}$
or $\bR_c$ implies infinite degeneracy of a given Landau energy level.
We can use $\bK$ to classify the eigenstates. 
However, since two components of $\bK_{\perp}$ do not commute,
$[K_x,K_y]=-ie_i B$, only one function of $K_x, K_y$ can be diagonalized
for stationary states. This means that the guiding center
of the particle can not be specified accurately.
If we use $K_x$ to classify the states,
then the wavefunction has the well-known form $e^{K_x x}\phi(y)$
\cite{Landau77}, where
the function $\phi(y)$ is centered at $y_c=-K_x/(e_iB)$,
as can be inferred from Eq.~(2.2). The Landau degeneracy in 
an area ${\cal A}_g=L_g^2$ is thus given by 
$(L_g/2\pi)\int\! dK_x=(L_g/2\pi)|K_{x,g}|={\cal A}_g
(|e_i|B/2\pi)$, where we have used $K_{x,g}=-e_iBL_g$.
On the other hand, if we choose to diagonalize $K_{\perp}^2=K_x^2+K_y^2$,
we obtain the Landau wavefunction $W_{nm}(\br_\perp)$
in cylindrical coordinates\cite{Landau77},
where $m$ is the ``orbital'' quantum number (denoted by $s$ 
in some references). For the ground Landau level this is
\be
W_{0m}(\br_\perp)\equiv 
W_{m}(\rho,\theta)={1\over (2\pi m!)^{1/2}\hat\rho}
\left({\rho\over \sqrt{2}\hat\rho}\right)^m 
\exp\left({\rho^2\over 4\hat\rho^2}\right)\exp(-im\theta).
\ee
The distance of the guiding center of the particle
from the origin is given by 
\be
\rho_m=(2m+1)^{1/2}\hat\rho,~~~~~m=0,1,2,\cdots,
\ee
where $\hat\rho$ is the cyclotron radius 
\be
\hat\rho=\left({\hbar c\over eB}\right)^{1/2}=
a_o\left({B_o\over B}\right)^{1/2}=b^{-1/2}~({\rm a.u.})
=2.57\times 10^{-10}B_{12}^{-1/2}~({\rm cm}).
\ee
The corresponding value of $K_{\perp}$ is given by 
$K_{\perp}^2=|e_i|B(2m+1)$. 
Note that $K_\perp^2$ assumes discrete values since $m$ is required to be an 
integer in order for the wavefunction to be single-valued. 
The degeneracy $m_g$ of the
Landau level in an area ${\cal A}_g=\pi R_g^2$ is then determined by
$\rho_{m_g} \simeq (2m_g)^{1/2}\hat\rho=R_g$, 
which again yields $m_g={\cal A}_g|e_i|B/(2\pi)$. 

We also note that $K_{\perp}^2$ is related to the $z$-angular
momentum $J_z$, as is evident from the $e^{-im\theta}$ factor in the
cylindrical wavefunction (Eq.~[2.3]). In general, we can show that
\be
J_z=xp_y-yp_x={1\over 2e_iB}(\bK_{\perp}^2-\bPi_{\perp}^2)
= (m-n) {|e_i|\over e_i},
\ee
where we have used $\bPi_{\perp}^2=|e_i|B(2n+1)$, and $n$ is
the quantum number for the Landau excitations.

\subsection{\bf Hydrogen Atom as a Two-Body Problem}

We now consider the electron-proton system.
It is easy to show that even with the Coulomb interaction
between the particles, the total pseudomomentum 
is a constant of motion
\be
\bK=\bK_1+\bK_2,
\ee
where the subscripts $1,2$ refer to electron (charge $-1$) and proton.
Moreover, unlike the
single particle case, here all components of $\bK$ commute.
Thus it is natural to separate the CM motion from the internal degree
of freedom, using the vector pseudomomentum $\bK$
as an explicit constant of motion (although we shall discuss
an alternative set of basic states in Sec.~IV). 
From Eq.~(2.2) we have
\be
\bR_K=\bR_{c1}-\bR_{c2}=-{\bK\times {\bf B}\over B^2}.
\ee
Thus we see that $\bK_{\perp}$ {\it is directly related to
the separation $\bR_K$ between the guiding center of the Landau orbit of the
electron and that of the proton}.

Consider the energy eigenstate with a fixed $\bK$.
Introduce the center-of-mass coordinate $\bR=(m_e\br_1+m_p\br_2)/(m_e+m_p)$ 
and the relative coordinate $\br=\br_1-\br_2$. 
Writing the two-body wavefunction as
\be
\Psi(\bR,\br)=\exp \left[i(\bK+{1\over 2}\bB\times\br)\cdot\bR\right]
\phi(\br),
\ee
so that $\Psi(\bR,\br)$ has a well-defined value of $\bK$, while
$\exp[i(\bB\times\br)\cdot\bR/2]$ is a gauge factor, the 
Schr\"odinger equation reduces to
\footnote{The spin terms of the electron and
the proton are not explictly
included. However, it should be understood that for the ground Landau state,
the zero-point Landau energy is exactly cancelled by
the spin energy. Also, the abnormal intrinsic magnetic moment of the proton 
is neglected, since it does not play a role in our analysis.}
\be
H\phi(\br)=(H_o+H')\phi(\br)=\cE\,\phi(\br),
\ee
with
\begin{eqnarray}
H_o &=& {K_z^2\over 2M}+{1\over 2\mu}\left(\bp+{1\over 2}\bB\times\br\right)^2
-{1\over m_p}\bB\cdot (\br\times\bp)-{1\over r},\\
H'  &=& {K_\perp^2\over 2M}+{1\over M}(\bK\times\bB)\cdot\br,
\end{eqnarray}
where $\bp=-i\partial/\partial\br$, 
$M=m_e+m_p\simeq m_p/m_e$ (a.u.), $\mu=m_em_p/M\simeq m_e$.
We will make no distinction between $M$ and $m_p$, nor $\mu$ and $m_e$ 
in our following calculations. Eqs.~(2.10)-(2.12) have already been 
derived in Refs.\cite{Lamb52,Gorkov68,Avron78,Herold81}.
Clearly, the CM motion is coupled to the internal motion 
through the second term in $H'$, which has the form of the potential
in a motion-induced electric field $(\bK/M)\times\bB$.
This term represents the so called ``motional Stark effect''
(although such a description is not exactly accurate, since
$\bK_{\perp}/M$ does not correspond to the CM velocity\cite{Johnson83}).
For small $K_\perp$, this effect can be considered by treating $H'$ as 
a perturbation (Sec.~III), but the eigenstates of $H_o$ can in principle
be used as a set of basic states for developing eigenstates of the full 
Hamiltonian for any value of $K_\perp$. However, the following 
transformed version of the Hamiltonian is more convenient for 
large $K_\perp$.

Motivated by the fact that $\bK_{\perp}$ measures the separation of the
guiding centers of the electron and the proton,
we can remove the ``Stark term''
by introducing a displaced coordinate $\br'=\br-\bR_K$, where
$\bR_K$ is given by Eq.~(2.8). After a gauge transformation, with
\be
\phi(\br)\rightarrow 
\exp\left(i{m_p-m_e\over 2M}\bK_\perp\cdot\br\right)\phi(\br'),
\ee
the Hamiltonian becomes
\be
H={K_z^2\over 2M}+{1\over 2\mu}\left(\bp'+{1\over 2}\bB\times\br'\right)^2
-{1\over m_p}\bB\cdot (\br'\times\bp')-{1\over |\br'+\bR_K|},
\ee
where $\bp'=-i\partial/\partial\br'$. This expression 
has been obtained in Refs.\cite{Avron78,Herold81}. 
We shall see in Sec.~III that this alternative form of 
the Hamiltonian is useful in the regime where $b$ is much larger than 
$b_{crit}$, defined in Eq.~(1.2). 

\section{\bf APPROXIMATE SOLUTIONS AND FITTING FORMULAE}

\subsection{\bf Zeroth Order Solutions}

We consider first the Hamiltonian formulation in terms of 
Eqs.~(2.11)-(2.12). For the zeroth order Hamiltonian $H_o$
the quantum numbers for the basic states are $K_z$, 
the number of nodes in the $z$-wavefunction 
$\nu$, the electron Landau level 
integer $n$ and the ``orbital'' quantum number $m=0,1,2,\cdots$. 
In this paper we only consider $b>>1$ and thus restrict ourselves to 
$n=0$. The energy eigenvalues of $H_o$ for the $\nu=0$ states
can be written as 
\be
\cE_{m}^{(0)}(K_z)={K_z^2\over 2M}+E_{m}+m{b\over M},
\ee
where $E_{m}$ is the energy of a bound electron in the fixed Coulomb potential
of an infinitely massive positive charge. The last term in Eq.~(3.1) 
for $m\ge 1$ represents Landau energy excitations for the 
proton, but $m$ is merely the ``orbital'' quantum number for the electron 
wavefunction and measures the relative $z-$angular momentum $J_z=-m$.
Thus, there is no separate quantum number for the proton in 
this formulation. The ``coupling'' between the electron quantum number
$m$ and the proton Landau excitation $mb/M$ in Eq.~(3.1) results from 
the conservation of total pseudomomentum.
The term $E_m$ has the form (e.g.,\cite{Ruderman74,Lai92})
\be
E_m\simeq -0.16 A\,l_m^2,~~~~l_m\equiv\ln{b\over 2m+1},
\ee
where $A$ is a coefficient which varies slowly with $b$ and $m$
(E.g., $A\simeq 1.01-1.3$ for $m=0-5$ when $B_{12}=1$, and 
$A\simeq 1.02-1.04$ for $m=0-5$ when $B_{12}=10$). In most formulae below, we 
replace $A$ by unity; the numerical values 
can be found in\cite{Lai92,Lai95}. The atom has a cigar-like shape, with 
size $\sim\rho_m$ (cf.~Eq.~[2.4]) perpendicular to the field and 
$L_z\sim l_m^{-1}$ along the field direction.
Note that the correction due to the reduced mass
$\mu$ could be easily incorporated by a simple scaling:
\be
E_m\simeq -0.16A{\mu\over m_e}\left[\ln
\left({b\over 2m+1}{m_e^2\over \mu^2}\right)\right]^2.
\ee
However, this is a small correction (of order $m_e/m_p$), and
will be neglected hereafter. 

Equations (3.2)-(3.3) refer to the ``tight-bound'' states for which 
the number of nodes $\nu$ of the $z$-wavefunction $f(z)$ of the electron 
is zero. For $\nu>0$, the energy eigenvalues are approximately given 
by\cite{Haines69}:
\be
E_{m\nu}=-{1\over 2\nu_1^2}\left(1-{4\rho_m\over\nu_1 a_o}\right),
~~~~\nu_1=1,2,3,\cdots
\ee
for the odd states ($\nu=2\nu_1-1$), and 
\be
E_{m\nu}=-{1\over 2\nu_1^2}\left[1-{2\over\nu_1\ln(a_o/\rho_m)}\right],
~~~\nu_1=1,2,3,\cdots
\ee
for the even states ($\nu=2\nu_1$).
The sizes of the wavefunctions are $\rho_m$ perpendicular to the field
and $L_z\sim \nu^2$ (a.u.) along the field. 
These states have much lower binding energies
compared to the ``tight-bound'' states. 

We now consider the energies and eigenstates of the atom for finite $K_\perp$. 
The two different Hamiltonian forms are discussed 
in Sec.~III.B-C, and the general approximate expressions for the energies
are then given in Sec.~III.D. We focus on the ``tight-bound'' states only,
since finite $K_\perp$ will make the weakly-bound $\nu>0$ states even
less bound (although in Sec.~V we will include an estimate of the 
statistical weight of these states in the partition function of 
bound atom). 

\subsection{\bf The Perturbation Hamiltonian Formalism}

For sufficiently small $K_\perp$, we can use standard perturbation 
theory to calculate the correction of energy $\cE$ due to $H'$
given by Eq.~(2.12) (see also\cite{Herold81,Vincke88,Pavlov93}).
Let $\bK_\perp$ be along the $y$-axis,
then the $r$-dependent part of $H'$ is $K_\perp bx/M$. 
We consider only $n=0$ and $\nu=0$, so the exact eigenstates of 
$H_o+H'$ are superpositions of the $H_o$-eigenstates with
$m=0,1,2,\cdots$. The only non-zero matrix elements of $x$ are of the form
\be
\langle W_{m}|x|W_{m+1}\rangle =\left({m+1\over 2}\right)^{1/2}
\hat\rho,
\ee
and the energy differences of adjacent $H_o$-eigenstates are 
approximately given by 
\be
\Delta \cE_m^{(0)}=\cE_{m+1}^{(0)}-\cE_m^{(0)}\simeq 
0.32\,l\,\ln\left({2m+3\over 2m+1}\right)+{b\over M},
\ee
where $l=l_0\equiv\ln b$ and the factor $0.32$ is an approximation 
to a slowly varying function of $m$ and $b$. We first consider
the ground state $m=0$.
Using $H'=K_\perp bx/M+K_\perp^2/(2M)$ and Eq.~(3.6) we
note that perturbation theory is justified if $K_\perp$ is much smaller
than a ``perturbation limit'' $K_{\perp p}$ defined as 
\be
K_{\perp p}={M\Delta\cE_0^{(0)}\over b^{1/2}}
=b^{1/2}\left(1+{Ml\over\xi b}\right)
\simeq b^{1/2}\left(1+{b_{crit}\over\xi b}\right),
\ee
where $\xi\sim 2.8$ is a slowly varying function of
$b$ (e.g., $\xi\simeq 2-3$ for $B_{12}=0.1-10^3$). 
For $K_\perp <<K_{\perp p}$, the energy $\cE_0^{(2)}$ to be added to 
$\cE_0^{(0)}$ in Eq.~(3.1) is given by second order perturbation theory
(plus a diagonal term) as
\be
\cE_0^{(2)}={K_{\perp}^2\over 2M}\left(1-{b\over M\Delta\cE_0^{(0)}}\right)
={K_\perp^2\over 2M_{\perp}},
\ee
where $M_\perp$ is the effective mass for the ``transverse motion'' 
of the atom,
\be
M_\perp= M\left(1+{\xi b\over lM}\right)
\simeq M\left(1+{\xi b\over b_{crit}}\right).
\ee
Thus the effective mass $M_\perp$ increases with increasing $b$. 

Similar calculations for the $m>0$ states yield 
$\cE^{(2)}_m=K_\perp^2/(2M_{\perp m})$, with
\be
1-{M\over M_{\perp m}}\simeq {b\over M}\left[{m+1\over b/M+0.16\,l_m^2
-0.16\,l_{m+1}^2}-{m\over b/M+0.16\,l_{m-1}^2-0.16\,l_m^2}\right].
\ee
A convenient (but approximate) expression for the effective mass 
$M_{\perp m}$ is given by
\be
M_{\perp m}\simeq M+\xi_m (2m+1){b\over l_m},
\ee
where $\xi_m$ is of the same order of magnitude as $\xi$, but different 
(by a factor of a few) for different $m$-states. 
The important feature in Eqs.~(3.11)-(3.12) is that the effective mass 
is larger for the higher-$m$ state. 

The quadratic form of the effective ``transverse kinetic energy''
in Eq.~(3.9) is valid only when it is much less than 
the ``perturbation limit'', reached when $K_\perp=K_{\perp p}$. 
Using Eq.~(3.8) and the approximation in Eq.~(3.10), this kinetic
energy limit becomes
\be
{K_{\perp p}^2\over 2M_\perp}=
{l\over 2\xi}\left(1+{Ml\over\xi b}\right)
\simeq 1.7\left(1+{b_{crit}\over\xi b}\right)~({\rm a.u.}).
\ee
For $b<<b_{crit}$ (even if $b>>1$) this limit is large 
compared with $1$ a.u., so that the quadratic 
perturbation energy (or transverse kinetic energy) in Eq.~(3.9)
is valid for the most important (low energy) states. 
Moreover, from Eq.~(3.10) the effective mass is close to the 
actually proton mass $M$. For superstrong fields,
$b>>b_{crit}$, on the other hand, the effective mass
(for $K_\perp<<K_{\perp p}$) is much larger than $M$
and the perturbation formalism already breaks down when the 
transverse kinetic energy is only of order $1$ a.u. 


At least for $B>>B_{crit}$, we have to consider values of $K_\perp$
large enough so that the perturbation treatment for the formalism in 
Eqs.~(2.10)-(2.12) is unsuitable. For any magnetic field strength
there are still eigenstates for arbitrarily large values of
$K_\perp$, but the transformed Hamiltonian in Eqs.~(2.13)-(2.14)
is now more suitable for calculating the energy. 

\subsection{\bf The ``Displaced Center'' Formalism}

As mentioned, Eq.~(2.14) gives an alternative formulation for the
Hamiltonian where $K_\perp^2/(2M)$ does not appear explicitly, but the 
displacement of the electron-proton guiding centers does with $R_K=K_\perp/b$
(in atomic units)\cite{Herold81,Potekhin94}. 
We again focus only on the Landau ground state
for the electron ($n=0$), but in principle we must include all
$m$-values (with proton Landau excitation energy $mb/M$) and 
mixing between these states. We consider first the approximation where we 
omit mixing, i.e., we use the diagonal matrix element of
$|\br'+\bR_K|^{-1}$ for a fixed $m$-value, and restrict 
ourselves to $\nu=0$, $K_z=0$ and $(2m+1)<< b$, so that 
$\rho_m^2=(2m+1)/b<<1$. We can estimate the size $L_z$ of the 
atom along the $z$-axis and the energy $\cE_m$ for two different regimes
of the values of $R_K$.

(i) For $R_K\lo L_z\lo 1$ (but not necessarily $R_K<\rho_m$) and 
with $L_z$ as a variational parameter we have
\be
\cE_m\sim {1\over L_z^2}-{1\over L_z}\ln {L_z^2\over
\rho_m^2+R_K^2}+{mb\over M},
\ee
where the first term is the kinetic energy along the $z$-axis, 
and the second term is
the potential energy of the electron. The logarithmic factor in Eq.~(3.14) 
comes from an integration over the ``cigar-shaped'' electron cloud 
in the displaced Coulomb potential $-1/|\br'+\bR_K|$. 
Minimizing $\cE_m$ with respect to $L_z$, we obtain
\be
L_z\sim \left(\ln {1\over\rho_m^2+R_K^2}\right)^{-1},
~~~~~\cE_m \sim -\left(\ln {1\over\rho_m^2+R_K^2}\right)^2+{mb\over M}.
\ee
The mixing between different $m$-states is unimportant when 
$b>> b_{crit}$. This can be seen from the order of
magnitude estimate of the off-diagonal matrix element between 
$m=0$ and $m=1$ states:
\be
\left\langle 0\left|{1\over |\br'+\bR_K|}\right| 1\right\rangle
\sim \left\langle 0\left|{\bR_K\cdot\br'\over r'^3}\right|1\right\rangle 
\sim {R_K\hat\rho \over L_z(\hat\rho^2+R_K^2)} \lo l,
\ee
as compared with $\Delta\cE_m\sim (l+b/M)\sim l(1+b/b_{crit})$. 
For $b\lo b_{crit}$ the mixing is non-negligible, especially 
when $R_K\sim \hat\rho<<1$; some results are given
in Ref.\cite{Potekhin94}. 
When $b>>b_{crit}$, the mixing can be neglected for all $R_K$. 

(ii) For $R_K\go 1$, the Coulomb logarithm in Eq.~(3.14) disappears, 
and we have (for $m=0$):
\be
\cE_0\sim {1\over L_z^2}-{1\over (L_z^2+R_K^2)^{1/2}}.
\ee
In the limit of $R_K>>1$, minimization of $\cE_0$ with respect to $L_z$
yields 
\be
L_z\sim R_K^{3/4}<<R_K, ~~~~\cE_0\sim -{1\over R_K}=-{b\over K_\perp}.
\ee
Thus for $K_\perp\go b$, the atom is very weakly bound 
($|\cE_0|\lo 1$). The limiting scaling relations in Eq.~(3.18) have
been identified in\cite{Vincke92,Potekhin94}. 

We can calculate the energy eigenvalue more accurately. 
Substituting $\phi(\br')=W_m(\br_\perp')f_m(z)$ into 
$H\phi(\br')=\cE\phi(\br')$
with $H$ given by Eq.~(2.14), and averaging over the transverse direction,
we obtain a one-dimensional Schr\"odinger equation
\be
-{1\over 2\mu}{d^2\over dz^2}f_m(z)+V_m(z,R_K)f_m(z)
=\cE_mf_m(z),
\ee
where the averaged potential is given by
\be
V_m(z,R_K)=-\left\langle W_m(\br'_{\perp})\left|{1\over |\br'+\bR_K|}
\right|W_m(\br'_{\perp})\right\rangle.
\ee
The function $V_m(z,R_K)$ can be evaluated using an integral 
representation (e.g.,\cite{Lai95})
\be
V_m(z,R_K)=-{1\over\hat\rho}\int_0^{\infty}\!\!dq
\exp\left(-{q^2\over 2}-q{|z|\over\hat\rho}\right)J_0(qR_K/\hat\rho)
L_m\left({q^2\over 2}\right),
\ee
where $J_0$ is the Bessel function of zeroth order and $L_m$ is the 
Laguerre polynormial of order $m$\cite{Abramowitz72}.
We solve for $E$ by integrating Eq.~(3.19) numerically from 
$z=\infty$ to $z=0$ subject to the boundary conditions
$df/dz=0$ at $z=0$ and $f\rightarrow 0$ as $z\rightarrow\infty$.
The energy eigenvalue $\cE_m$ as a function of $R_K$ are shown
in Figure 1 for $B_{12}=10$ and $100$. 
Similar numerical results have also been obtained in Ref.\cite{Potekhin94}. 
For $R_K\lo 1$,
we can fit the energy to a form similar to Eq.~(3.15). The total energy
of the $m=0$ state is then given by
\be
\cE_0(K_z,K_{\perp})\simeq {K_z^2\over 2M}
-0.16\left(\ln {1\over \hat\rho^2+CR_K^2}\right)^2,
\ee
and the atomic size $L_z\sim 1/|\ln (\hat\rho^2+CR_K^2)|$.
Equation (3.22) reduces to Eq.~(3.2) for $K_\perp=0$.
From the numerical results we find $C\simeq 0.8$. 
For $R_K\go 1$, the binding energy of the atom is much smaller, and
Eq.~(3.22) should be replaced by the order of magnitude relation (3.18),
while the actual numerical values of the energies are not important 
in practice (see Sec.~V). 

In the $B>>B_{crit}$ regime, the maximum value $K_{\perp p}$ for the 
perturbation treatment of Sec.~III.B to be valid becomes 
$K_{\perp p}\sim b^{1/2}$ (cf.~Eq.~[3.8]). Thus for $K_\perp<<b^{1/2}$,
Eq.~(3.22) should be consistent with Eq.~(3.9). Indeed, 
when $R_K << \hat\rho$, Eq.~(3.22) reduces to 
\be
\cE_0(K_z,K_{\perp})\simeq {K_z^2\over 2M}
-0.16\,l^2+0.32\,C{l\over b}K_{\perp}^2. 
\ee
The dependence on $K_\perp$ is again quadratic, and the 
corresponding effective transverse mass is 
$M_{\perp}\simeq b/(2ACl)\simeq 2b/l\simeq \xi b/l$, in agreement 
with Eq.~(3.10). 

The $m>0$ states can be similarly calculated using Eqs.~(3.19)-(3.21).
Some numerical results are again shown in Figure 1.
The energy can be expressed approximately as 
\be
\cE_m(K_z,K_{\perp})\simeq {K_z^2\over 2M}+
m{b\over M}-0.16\left(\ln {1\over \rho_m^2+C_mR_K^2}\right)^2.
\ee
From the numerical results we again obtain $C_m\sim 1$. 

Comparison with the numerical results of Potekhin\cite{Potekhin94},
who included the mixing of different $m-$states, indicates that 
Eqs.~(3.22) and (3.24) are accurate to within $\sim 30\%$ in the
relevant regime of $K_\perp$ ($\lo b$) when $b\go b_{crit}$. The 
agreement becomes better as $b$ increases. For smaller $b$, however, 
the perturbative results of Sec.~3.B should be adequate (see
Sec.~3.D). 

Finally, if we consider sufficiently strong magnetic field 
so that not only $b/M\go l$ (or $b\go b_{crit}$) but also $b/M\go l^2$ is 
satisfied, then Eq.~(3.24) implies that all the $m>0$ states are unbound, 
as has already been noted in Ref.\cite{Herold81}. However, this does 
{\it not} mean that there is no other bound state except a single 
non-degenerate $m=0$ state. Indeed, Eq.~(3.22) indicates 
that there are many 
states for which the guiding centers of proton and electron are separated 
by a small $R_K$, and these states have similar energies compared to 
the ground state ($m=0,K_\perp=0$). 
In the pseudomomentum scheme discussed here, these 
states occupy a continuum 
$K_\perp$-space. As we shall see in Sec.~IV, these closely-packed 
energy levels can be made discrete if we use a different set of eigenstates.

\subsection{\bf General Fitting Formulae}

Consider for the moment the cases with $K_\perp<<b$, i.e.,
$R_K<<1$, but with no other restrictions on $K_\perp$ or $b$. 
In Sec.~III.B-C, we have obtained reasonably accurate ground-state 
($m=0$) energy of a hydrogen atom in two limiting regimes:
(i) for $B<<B_{crit}$, where Eqs.~(3.9)-(3.10) are applicable 
up to adequately large values of $K_\perp$, and 
(ii) for $B>>B_{crit}$, where the energy is given by Eq.~(3.22).
We write the total ground state energy in the form
\be
\cE_0(K_z,K_{\perp})\simeq {K_z^2\over 2M}-0.16\,l^2+E_\perp(K_\perp),
\ee
and want to find a general fitting formula for the 
``transverse kinetic energy'' $E_\perp(K_\perp)$ with 
$K_\perp<<b$, but $K_\perp/b^{1/2}=R_K/\hat\rho$
otherwise arbitrary. With the inequality $R_K<<1$, the 
second term in Eq.~(3.22) can be approximated by 
$-0.16\,l^2+0.32\,l\,\ln(1+CR_K^2/\hat\rho^2)$. 
We propose the following fitting formula
\be
E_\perp(K_\perp)\simeq {\tau\over 2M_\perp}\ln\left(1+{K_\perp^2\over\tau}
\right),~~~~~\tau\simeq 0.64\,\xi K_{\perp p}^2,
\ee
where $M_\perp$ is given by Eq.~(3.10). 
Recall that the parameter $\xi\simeq 2-3$ for $B_{12}=0.1-10^3$, and 
a typical number to use is $\xi=2.8$. 
This formula reduces to Eq.~(3.9) when $K_\perp<<K_{\perp p}$
and to the above approximation of Eq.~(3.22) when $B>>B_{crit}$.
We expect Eq.~(3.26) to be accurate to within $30\%$ for $K_\perp<<b$.
In the regime $b>>b_{crit}$ and $b<<K_\perp^2<<b^2$
(i.e., $\hat\rho<<R_K<<1$), Eq.~(3.26) reduces to 
$E_\perp\simeq 3.1\ln(K_\perp^2/b)~({\rm a.u.})>>3~$(a.u.), 
which is large compared with the thermal energy (temperature) 
$T$ for the astrophysical applications of interest. For
evaluating the integral over $K_\perp$ extending from zero
to infinity in the atomic partition function (Sec.~V), we
shall advocate using Eq.~(3.26) for all $K_\perp$ even though
this expression tends to infinity, 
\footnote{A slightly more general fitting formula which closely
resembles Eq.~(3.22) is given by 
$$\cE_0(K_z,K_\perp)\simeq {K_z^2\over 2M}-0.16\left[\ln
{b\over (1+K_\perp^2/\tau)^\Gamma}\right]^2,$$
with $\Gamma=1+Ml/(\xi b)=\tau/(0.64lM_\perp)$.
This expression can be applied to $K_\perp\lo b$ and gives the 
correct limiting result for $K_\perp>>b$, but it is 
not convenient to use in practice. For the applications discussed 
in Sec.~V, Eqs.~(3.25)-(3.26) are adequate.}
whereas the correct expression should approach the finite limit 
$0.16\,l^2$ for $K_\perp>>b$. 
The difference is appreciable only
where $E_\perp>>T$ (so that the Boltzman factor $e^{-E_\perp/T}$
is very small) and our prescription amounts to ``cutting off'' 
the integral, i.e., omitting the states with $R_K>>1$ from the
integral. This omission is advantageous, since these
states should be treated together with the ionized states, 
and both turn out to be unimportant (Sec.~V). 


Our fitting formulae for the energies of the $m>0$ states are less accurate. 
In the small-$K_\perp$ limit, the $K_\perp$-dependent term in Eq.~(3.24) 
reduces to the quadratic form $K_\perp^2/(2M_{\perp m})$, with 
the effective mass given by Eq.~(3.12). Similar to Eqs.~(3.25)-(3.26) 
we fit $\cE_m$ to the analytical form:
\be
\cE_{m}(K_z,K_{\perp})\simeq {K_z^2\over 2M}+m{b\over M}
-0.16\,l_m^2+E_{\perp m}(K_\perp),
\ee
where
\be
E_{\perp m}(K_\perp)\simeq {\tau_m\over 2M_{\perp m}}
\ln\left(1+{K_\perp^2\over\tau_m}\right),~~~~~
\tau_m\simeq 0.64\,\xi_m (2m+1)b\left[1+{Ml_m\over\xi_m (2m+1)b}\right]^2,
\ee
so that $E_{\perp 0}=E_\perp$, $\tau_0=\tau$.
Although $\xi_m$ can vary by a factor of a few for different values 
of $B$ and $m$, equation (3.28) has the correct approximate
functional dependence for a wide range of $B$ and $K_\perp$.

As noted in the introduction, one cannot construct a wave packet
to mimic the classical behavior of a moving atom when both electron and 
proton are confined to the ground Landau level.
Therefore when $B>>B_{crit}>>B_o$, the notion of ``motion across the 
magnetic field'' does not have a unique meaning. Nevertheless,
one can ask two questions about the transverse pseudomomentum: 
(i) Is there an upper limit to $K_\perp$ beyond which there is no 
bound state? (ii) When a bound state exists, what is the value of
${\bf V}_\perp\equiv \partial{\cal E}/\partial {\bf K}_\perp$,
the analog to the classical center-of-mass
transverse velocity of the atom? From our discussion in this section, 
we have seen that there exist bound states for all values of $B$ and
$K_\perp$, although the states with large $K_\perp$ are very weakly
bound. From Eq.~(3.26), we have 
$${\bf V}_\perp \simeq {{\bf K}_\perp\over M_\perp}\left(1+{K_\perp^2
\over\tau}\right)^{-1}.\eqno(3.29)$$
(Note that for $K_\perp>>b$, Eq.~[3.18] should be used and we have
$V_\perp\sim b/K_\perp^2$ instead). 
Clearly, in general $V_\perp$ is smaller
than its field-free counterpart: $V_\perp\simeq K_\perp/M_\perp$ 
(so that the effective mass description is valid) only for
$K_\perp<K_{\perp p}$, and $M_\perp\simeq M$ (classical
behavior) only for $b<<b_{crit}$. As $K_\perp\rightarrow\infty$,
the center-of-mass velocity approaches zero. For a given magnetic
field strength, the maximum $V_{\perp}$ is given by
$$V_{\perp max}\simeq {\tau^{1/2}\over 2M_\perp}\sim 
{{K_\perp p}\over M_\perp}\sim l\,b^{-1/2},\eqno(3.30)$$
which occurs when $K_\perp\simeq \tau^{1/2}\sim  K_{\perp p}$.
For $b\go b_{crit}$, the states with $K_\perp>>K_{\perp p}$
have not only small velocity but also large electron-proton
separation $R_K>>\hat\rho$. Nevertheless, these states are quite distinct 
from an electron-proton pair with separation $R_K$ in fully 
ionized hydrogen, because the relative $z$-coordinate
satisfies a bound-state wavefunction (localized, although 
with large scale-length and small binding energy). 
The partition function for these states is smaller than 
that for the ionized component, because the sum over the 
relative momentum in the $z$-direction,
$k_z$, is absent. 

\section{\bf ALTERNATIVE APPROACH TO THE TWO-BODY PROBLEM}


The basic states used in Sec.~III for the e-p two-body problem
are explicit eigenstates of the transverse pseudomomentum $\bK_\perp$.
For $K_\perp=0$, this formulation has the advantages that the 
electron's ``orbital'' number $m$ is a good quantum number, 
the wavefunction can be related to that for a fixed positive 
charge and the energy is given explicitly by Eq.~(3.1). The 
$K_\perp=0$ states with different values of $m$ are 
orthogonal to each other and could be used to satisfy the 
Pauli principle for the electrons in a hydrogen 
molecule\cite{Lai92,Lai95} or a multi-electron atom\cite{Neuhauser87}.
The last term $mb/M=m(b/b_{crit})\ln b_{crit}$
in Eq.~(3.1) is unimportant when $b<<b_{crit}$,
but it is very large when $b>>b_{crit}$ (and $m\ge 1$).
The simplest wavefunction for the H$_2$-ground state in 
this formulation would use one $m=0$ and one $m=1$ 
electron\cite{Lai92}, and the energy would include the positive term 
$(b/b_{crit})\ln b_{crit}$. One therefore might conclude 
that H$_2$ has a positive energy relative to two H atoms
when $b>>b_{crit}$. In a future paper\cite{Lai95} we will 
show that this is not the case. In a molecule, only 
the total pseudomomentum is conserved, not that for individual 
atoms. Thus for $b>>b_{crit}$ one would have to perform
an integral over $K_\perp$ of individual atoms to form a
molecular eigenstate (Similar situation exists for multi-electron 
atoms). To our knowledge this complicated task has not yet been 
carried out.  We propose here an alternative set of
basic states which are not eigenstates of the pseudomomentum.
This formulation is more complex for single H atom, but can be 
generalized more easily to molecule. Moreover, this formulation
makes the definition of eigenstates clearer (unlike the 
pseudomomentum approach, where $m$ is a good quantum only for 
$K_\perp=0$), and therefore offers an intuitive understanding 
of the degeneracy of states.

Let $\br_1$, $\br_2$ be the position vectors of electron and proton
in a H atom. We introduce coordinates $Z=(m_ez_1+m_pz_2)/M$ 
and $z=z_1-z_2$, but stick to $\br_{1\perp}$
and $\br_{2\perp}$. In this mixed coordinate system, the Hamiltonian for
the electron-proton pair becomes $H=H_o+V$, with (cf.~footnote 1):
\be
H_o={K_z^2\over 2M}+{p_z^2\over 2\mu}+\sum_{i=e,p}{1\over 2m_i}
(\bp_{i\perp}-e_i\bA_i)^2,
\ee
where $\bA_i=\bA(\br_i)=\bB\times\br_{i\perp}/2$, 
and the interaction potential is
\be
V=-{1\over [z^2+(\br_{1\perp}-\br_{2\perp})^2]^{1/2}}.
\ee
We set $K_z=0$ without loss of generalality. 

Consider the transverse part of the wavefunction for a bound state. 
The Landau wavefunctions of the electron and proton form a complete set
\be
W_{n_1m_1}(\br_{1\perp})W_{n_2m_2}^{*}(\br_{2\perp}),~~~~n_1,n_2,m_1,m_2=0,1,
2,\cdots
\ee
($n_1,n_2$ specify the Landau excitations, $m_1,m_2$ are the 
``orbital'' quantum numbers as in 
Eqs.~[2.3]-[2.4]). In general, an eigenstate of $H$ can be constructed as 
\be
\Psi(z,\br_{1\perp},\br_{2\perp})=
\sum_{n_1,m_1,n_2,m_2}W_{n_1m_1}(\br_{1\perp})W_{n_2m_2}^{*}(\br_{2\perp})
f_{n_1m_1n_2m_2}(z),
\ee
where we have restricted ourselves to $z$-wavefunctions without
a node ($\nu=0$). 
Substituting Eq.~(4.4) into the Schr\"odinger equation $(H_o+V)\Psi=\cE\Psi$
and using the orthogonal relations for the functions $W$ to average over the 
transverse direction, we obtain a set of coupled differential 
equations from which the functions $f(z)$'s and the eigenvalue $\cE$
can be calculated, at least in principle. 

This set of equations is greatly simplified as a result of the conservation of 
total $z$-angular momentum $J_z$. Form Eq.~(2.6), we have
\be
J_z=-(m_1-n_1)+(m_2-n_2).
\ee
Indeed, since the basis function 
$W_{n_1m_1}(\br_{1\perp})W_{n_2m_2}^{*}(\br_{2\perp})\propto 
e^{-i(m_1-n_1)\theta_1+i(m_2-n_2)\theta_2}$, 
while the interaction potential $V$
depend only on $(\theta_1-\theta_2)$, we readily see that only the states
with the same $J_z$ are coupled. 

We shall use the formalism of this section only when 
$B>>B_{crit}$, in which case the Landau energy of the proton 
is large compared to the atomic Coulomb energy, so that 
both $n_1$ and $n_2$ become good quantum numbers. For 
astrophysical applications we are then interested in the 
ground Landau levels, $n_1=n_2=0$. 
The energy eigenstate with a fixed $z$-angular momentum
$J_z=m_t=m_2-m_1$ can be constructed as
\be
\Psi^{(m_t)}(z,\br_{1\perp},\br_{2\perp})=
\sum_{m_1}W_{m_1}(\br_{1\perp})W_{m_t+m_1}^{*}(\br_{2\perp})
f^{(m_t)}_{m_1}(z).
\ee
The equations for the functions $f$'s are then given by
\be
-{1\over 2\mu}{d^2\over dz^2}f^{(m_t)}_{m_1}(z)
-\sum_{m_1'}G^{(m_t)}_{m_1m_1'}(z)
f^{(m_t)}_{m_1'}(z)=\cE^{(m_t)}f^{(m_t)}_{m_1}(z),
~~~~m_1=0,1,2,\cdots
\ee
where
\be
G^{(m_t)}_{m_1m_1'}(z)=
\langle W_{m_1}(\br_{1\perp})W_{m_t+m_1}^{*}(\br_{2\perp})
|{1\over |\br_1-\br_2|}
|W_{m_1'}(\br_{1\perp})W_{m_t+m_1'}^{*}(\br_{2\perp})\rangle.
\ee
The set of equations (4.7)-(4.8) 
essentially forms a differential-integral equation system
(the sum over $m_1$ can be considered as an integration). 
Some mathematical formulae for evaluating the function 
$G_{m_1m_1'}^{(m_t)}(z)$ in Eq.~(4.8) are given in Appendix A. 


Since the states with different $m_t$ are orthogonal, 
we can use the variational principle for each value of $m_t$ 
separately to find a rigorous upper bound to $E^{(m_t)}$.
To this end, we choose as a simple trial wavefunction 
the first term in Eq.~(4.6), 
i.e., we include only $m_1=0,~m_2=m_t$. Equation (4.7) then reduces
to a single differential equation which is straightforward to solve
numerically. For $m_t=0-4$, we find that the upper bound can 
be written in the form: 
\be
\cE^{(m_t)} < -0.16\,A\left(\ln {b/2.3\over C'm_t+1}\right)^2, 
\ee
with $C'\simeq 0.9-1.1$ (depending on the values of $m_t$), and 
$A$ is a coefficient close to unity (as in Eq.~[3.2]). 
This expression is equivalent to Eq.~(3.22), with $m_t^{1/2}\hat\rho$ 
playing the role of $R_K$, but here $m_t$ is an integer.  
The form in Eq.~(4.9) is expected since $m_t=m_2-m_1$ measures 
the difference between the distances to the origin 
of the guiding center of the electron and that of the proton.
Although the $m_t\ge 1$ state is not an exact $K_\perp$-eigenstate,
in a qualitative sense the separation between electron and 
proton increases with increasing $m_t$. Thus we expect 
\be
G^{(m_t)}(z) \sim {1\over [z^2+(C'm_t+1)\hat\rho^2]^{1/2}}, 
\ee
which then naturally leads to the form in Eq.~(4.9). 
The decrease from $b$ in Eq.~(3.2) to $b/2.3$ in Eq.~(4.9) arises
from the fact that both the electron and proton wavefunctions 
have finite spread around the same axis (unlike the usual 
pseudomomentum approach, where the relative coordinate is used).

The actual energy eigenvalue $\cE^{(m_t)}$ can be obtained by 
solving iteratively the series of equations in (4.7) 
using the standard shooting algorithm\cite{Press87}. 
We have carried out the calculations for the $m_t=0$ 
and $m_t=1$ states. Our numerical results are given 
in Figure 2 for three different values of field
strength $B_{12}=100,~1000$ and $5000$. 
Typically, more than $10$ terms in the sum of Eq.~(4.6)
are needed in order to attain convergence of the energy
to within $\lo 1\%$. 
We find that the ground-state energy eigenvalue $\cE^{(0)}$ 
agrees with the standard value (Eq.~[3.2] with $m=0$) 
in the limit of $B>>B_{crit}$. This is expected
because the $m_t=0$ ground state is also a $K_\perp=0$
eigenstate (Note that $\bK_{\perp 1}=-\bK_{\perp 2}$ implies
$m_1=m_2$ and hence $m_t=0$).  
Also, we see that as $B$ increases, the upper bound 
given by Eq.~(4.9) becomes asymptotically closer to 
the actual energy $\cE^{(m_t)}$. For general $m_t$, 
we cannot give a rigorous lower bound to $E^{(m_t)}$, but it 
is approximately given by Eq.~(4.9) with $(b/2.3)$ replaced 
by $b$, so that the fractional uncertainty decreases 
as $B$ increases. Note that although the contribution of the
of $m_1\ge 1$ terms in Eq.~(4.6) to the correct energy
becomes smaller as $B$ increases, the contribution
to the correct wavefunction is always non-negligible. 


It is instructive to consider the degeneracy of an energy 
eigenstate. Without the Coulomb interaction between electron 
and proton, there is a double Landau degeneracy in $m_1$ and $m_2$. 
When the Coulomb interaction is included, the degeneracy in 
$m_t=m_2-m_1$ is removed, but a single degeneracy remains in 
$m_1$, i.e., for a given $m_t$, the eigenfunction corresponding 
to $\cE^{(m_t)}$ is not unique: the eigenstate in Eq.~(4.6) with 
real $f_{m_1}$'s 
is presumably the state where the proton is centered at 
the origin; but there must be other states with the same energy, 
centered at different positions. 
We can demonstrate this degeneracy explicitly as follows. 
A Landau wavefunction centered at the origin of the coordinate
can be expanded in terms of wavefunctions centered at some 
point $\br_o=\br_{o\perp}$ as
\be
W_{nm}(\br_\perp)=\sum_{m'}e^{i(m'-m)\theta_o} 
I_{n+m,n+m'}\!\left({r_o^2\over 2\hat\rho^2}\right)
e^{i\bK_o\cdot\br_\perp/2}W_{nm'}(\br_\perp-\br_o),
\ee
where $\br_o\equiv x_o+iy_o=r_o e^{i\theta_o}$, $\bK_o$ 
is given by $\br_o=-\bK_o\times\bB/B^2$ (see Eq.~[2.2]),
$I_{n-m,n-m'}$ is the polynomial as defined in\cite{Virtamo75}, 
and $e^{i\bK_o\cdot\br_\perp/2}$ is the gauge factor.
We consider only the ground Landau level $n=0$, and write
\be
W_m(\br_\perp)= e^{i\bK_o\cdot\br_\perp/2}\sum_{m'}C_m^{m'}
W_{m'}(\br_\perp-\br_o),
\ee
where $C_m^{m'}$ is a (complex) function of $m,~m'$ and $\br_o$.
The energy eigenstate in Eq.~(4.6) can then be written as
\be
\Psi^{(m_t)}(\br_1,\br_2)=e^{i\bK_o\cdot(\br_{1\perp}-\br_{2\perp})/2}
\sum_{m_t'}\left[\sum_{m_1'}W_{m_1'}(\br_{1\perp}-\br_o)
W_{m_1'+m_t'}^{*}(\br_{2\perp}-\br_o) f_{m_1'}^{(m_t m_t')}(z)
\right],
\ee
where we have defined
\be
f_{m_1'}^{(m_t m_t')}(z)\equiv\sum_{m_1} C_{m_1}^{m_1'}
C_{m_1+m_t}^{m_1'+m_t'} f_{m_1}^{(m_t)}(z).
\ee
However, the term inside $[\cdots]$ in Eq.~(4.13) is exactly 
an energy eigenstate $\Psi^{(m_t')}(\br_1-\br_o,\br_2-\br_o)$
with $J_z'=m_t'$ based on the coordinate system centered at 
$\br_o$. Since the state represented by the left-hand-side 
of Eq.~(4.13) has a definite energy $\cE^{(m_t)}$, while
states with different $m_t'$ have different energies, we must 
have $m_t'=m_t$ in Eq.~(4.13). Thus
\be
\Psi_i^{(m_t)}(\br_1,\br_2)=e^{i\bK_o\cdot(\br_{1\perp}-
\br_{2\perp})/2}\Psi_j^{(m_t)}(\br_1-\br_o,\br_2-\br_o),
\ee
where we have added the subscripts ``i,~j'' to indicate that 
there are many states associated with a given $m_t$, i.e.,
the states with the same energy $\cE^{(m_t)}$ is not unique. 
Clearly, the degeneracy (per unit area) for a given $m_t$ is
$B/(2\pi)$, i.e., a single Landau degeneracy (see the discussion 
following Eq.~[2.5]). 

The above discussions demonstrate that there is a discrete set of 
states with $m_t=1,2,3,\cdots$, 
all having similar energies as the ground state ($m_t=0$),
and do not have any positive contribution $m_tb/M$ in their
energies. This has important consequences for the binding of 
hydrogen molecules in the $B>>B_{crit}$ regime. 
In a forthcoming paper\cite{Lai95} 
we shall use one $m_t=0$ and $m_t=1$ atom to construct the 
wavefunction for the H$_2$ ground state, which also does not 
involve any Landau excitation of the proton. 


\section{\bf IONIZATION-RECOMBINATION EQUILIBRIUM}

\subsection{\bf Overview}

We now consider the ionization-recombination equilibrium of hydrogen atoms,
e$+$p$\Leftrightarrow$ H, given by the generalized Saha equation in the 
presence of a strong magnetic field. Previous treatments of this 
problem (e.g.,\cite{Khersonskii87,Miller92,Ventura92,Pavlov94}) 
either ignored the coupling between the center-of-mass motion and the 
internal atomic structure, or did not have available our generalized 
formula for the ``transverse kinetic energy'' as a function of 
the pseudomomentum $K_\perp$. 

Let $T$ be the gas temperature in atomic units (about $3.16\times 10^5$ K),
so that the Boltzmann constant $k_B$ is set equal to unity, and $n_g$ 
be the number density (also in a.u.) of protons (either free
or bound) in the gas. We write $V_g=\pi R_g^3\equiv 1/n_g$ and 
$\cA_g=\pi R_g^2$, so that a ``Wigner-Seitz cylinder'' of
radius and length $R_g$ contains one proton on the average. Some of
the partition function integrals can be simplified if the density 
and temperature satisfy three inequalities: (i) The density is low
in the sense that $R_g=(\pi n_g)^{-1/3}$ is much larger than 
the largest dimension (i.e., the $z$-dimension) $L_z\sim l^{-1}$ of 
the ground-state atom; (ii) The temperature is much smaller than
the ground-state binding energy $|E_0|=|E(H)|\simeq 0.16\,l^2>>1$;
(iii) The Coulomb attraction between a proton and an electron 
at typical separation $R_g$ is of order $R_g^{-1}$; we assume 
$T>>R_g^{-1}$ so that the ``imperfect gas corrections'' are small.

The Saha equation involves the bound-atom partition function $Z(H)$ 
compared with the product $Z(e)Z(p)$ of the two free-particle
partition functions. Each of these systems has six discrete or
``nominally continuous'' quantum numbers. Two of these for the 
free e-p system refer to the $z$-motion, which can be 
represented by $K_z$, the center-of-mass $z$-momentum and $k_z$, the 
relative $z$-momentum. For the bound system (H atom), 
the $K_z$-partition function is identical, but instead of $k_z$ we have
the quantum number $\nu$. In both systems the electron Landau 
excitations have the same quantum number $n$ and energy $nb$, 
so that the $n$-partition functions are the same
(In practice, $b>>T$, so we only need to consider the ground Landau state 
of electron and the $n$-partition function is essentially unity). 
For the bound system,
the remaining three quantum numbers are $m$ and the two Cartesian 
components of the transverse pseudomomentum $\bK_\perp$. For the free e-p
system the three quantum numbers are $n_2$, the proton Landau
level integer, and the two transverse parameters $|K_{\perp 1}|$
and $|K_{\perp 2}|$. 

\subsection{\bf The Bound-State Partition Function of the H Atom}

We first consider only the ground state of the H atom, with $m=\nu=0$.
Using Eq.~(3.25), the canonical partition function in the
volume $V_g=R_g\cA_g$ of a ``Wigner-Seitz cylinder'' can be written as 
\be
Z(H) \simeq R_g\left({MT\over 2\pi}\right)^{1/2}
\exp\left({|E(H)|\over T}\right)Z_\perp,
\ee
where the factor $R_g(MT/2\pi)^{1/2}$ results from the free center-of-mass 
$z$-motion and $|E(H)|=|E_0|\simeq 0.16\,l^2$ is the ground-state 
binding energy. The partition function $Z_\perp$ associated 
with the ``transverse motion'' of the atom has the form
\be
Z_\perp={\cA_g\over (2\pi)^2}\int_0^{K_{\perp max}}\!\!
2\pi K_{\perp}dK_{\perp}\exp\left[-{E_\perp(K_\perp)\over T}\right],
\ee
where $E_\perp(K_\perp)$ is the generalized ``transverse kinetic
energy''. 
The upper limit $K_{\perp max}$ of the integral in Eq.~(5.2) 
is determined by the condition $R_K\lo R_g$ so that the pressure-ionized 
states are excluded in the bound state partition function.
It thus has the density-dependent form $K_{\perp max}\sim bR_g$. 
As discussed in Sec.~III.D, $E_\perp$ is well approximated by 
Eq.~(3.26) for $K_\perp$ up to $\sim b$. However, for $K_\perp>>b$
(or $R_K>>1$) the correct expression for $E_\perp$ is close to 
$0.16\,l^2$ (almost independent of $K_\perp$), whereas the 
approximate $E_\perp$ in Eq.~(3.26) increases with $K_\perp$.
At very low density ($R_K>>1$), there are highly excited
states with $K_\perp$ between $b$ and $K_{\perp max}\sim bR_g$, 
whose contribution to $Z_\perp$ is proportional to 
$\cA_g b^2n_g^{-2/3}\exp(-0.16l^2/T)$.
However, these states cover only a narrow range
of binding energies (of order $R_g^{-1}<<T$) and can be neglected 
compared with the ionized components, in view of the inequality
$0.16\,l^2>>T$. We can therefore omit these states entirely,
but also make only a small error if we merely replace $E_\perp$ by 
the approximation in Eq.~(3.26), which is an overestimate for
$b\lo K_\perp\lo bR_g$. This approximation has the
advantage that the extension of the integral in Eq.~(5.2) 
to infinity is not only finite but also small. We therefore
get a convenient expression for $Z_\perp$ by extending 
the integration to $K_{\perp max}\rightarrow\infty$. We have
\be
Z_{\perp}\simeq {\cA_g\over 2\pi}\int_0^\infty\!\!
K_{\perp}dK_{\perp}\exp\left[-{\tau\over 2M_\perp T}
\ln\left(1+{K_{\perp}^2\over\tau}\right)\right]
=A_g{M_\perp'T\over 2\pi},
\ee
with
\be
M_\perp'=M_\perp\left(1-{2M_\perp T\over\tau}\right)^{-1}.
\ee
Thus $Z_\perp$ is proportional to $M_\perp'$, which is larger than
$M_\perp$ (or equal to it when $\tau/(2M_\perp T)>>1$, so that the
effective mass approximation is valid throughout the regime of interest), 
and $M_\perp$ is larger than the actual mass $M$. While the transverse
motion is ``slowed down'' in the sense that 
$\partial E_\perp(K_\perp)/\partial K_\perp$ is smaller than the 
zero-field result, the $K_\perp\neq 0$ states still exist and their 
statistical weight $\propto M_\perp'$ is actually increased over the
zero-field result by the strong magnetic field. 

We now consider the internal partition functions associated with the 
$\nu>0$ and $m>0$ excited states, i.e., we write the total bound-state
partition function as
\be 
Z(H)=V_g\left({MT\over 2\pi}\right)^{1/2}{M_\perp'T\over 2\pi}
\exp\left({|E(H)|\over T}\right)z_\nu(H) z_m(H).
\ee
Start with the quantum number $\nu$. The internal 
partition function relative to the ground state is 
\be
z_\nu(H)\simeq 1+\exp\left(-{|E_0|\over T}\right)\sum_{\nu=1}^{\nu_{max}}
2\exp\left({|E_\nu|\over T}\right).
\ee
Here $E_\nu$ is given by Eqs.~(3.4)-(3.5) but can be approximated by
$E_\nu=-1/(2\nu^2)$, and the factor of $2$ comes from 
the near-degeneracy of the even and odd states. 
The sum in Eq.~(5.6) is thus similar to
that for a field-free atom except that the usual weight-factor $2\nu^2$
is missing as the atom is one-dimensional. Since $|E_0|-|E_\nu|>>T$
for the temperature regime of interest, the 
individual term in the sum is very small, but the $z_\nu(H)$ would diverge if 
$\nu_{max}$ were allowed to become infinite. As for the 
high Rydberg states in the field-free H atom, the divergence is 
avoided by including only states which fit inside the Wigner-Seitz
cylinder. The size of the $\nu\ge 1$ state is $L_z\sim\nu^2$ (a.u.), so 
we should choose $\nu_{max}\sim R_g^{1/2}$ with energy 
$E_{\nu_{max}}\simeq -1/(2R_g)$, i.e., we omit states with such 
extended wavefunctions that they would be pressure-ionized. 
With this prescription the sum in Eq.~(5.6) becomes finite but 
density-dependent, i.e., it increases as $n_g^{-1/6}$ with decreasing 
$n_g$. Making use of the inequality $T>>R_g^{-1}\simeq 2|E_{\nu_{max}}|$,
we split the sum into two parts: one extends from $\nu=1$ to $\nu_T$, where 
$|E_{\nu_T}|=T$; the other from $\nu_T$ to $\nu_{max}$
(which exceeds $\nu_T$ because of the inequality).
The first part contains only a few terms, each with $|E_0|-|E_\nu|>>T$, 
so they can be neglected (compared with unity, coming from the 
$\nu=0$ term). The second part represents the highly excited states and
would contribute approximately $2\exp(-|E_0|/T)(\pi n_g)^{-1/6}$, 
which could be large for very small density $n_g$. 
However, these states should be considered 
separately and compared with the ionized components: These states have 
negative energies of order $(-1/R_g)$, while the ionized components 
have positive energies of order $T$. Because of the inequality 
$T>>R_g^{-1}$, we neglect these states entirely in the rest of the paper
and set $z_\nu(H)\simeq 1$. 

The contribution of the $m>0$ states to the bound-state partition 
function can be considered in the same manner as that of the $m=0$ 
state discussed before. The internal partition function 
associated with the $m$-states is given by
\footnote{The factor $(1+e^{-b/MT})$ in Eq.~(5.7) results
from the proton spin term, which is not explicitly included
in our calculations (see footnote 1). Note that when the proton 
spin term (but not the abnormal magnetic moment) is taken into 
account in Eq.~(2.11) or (2.13), the energy of the $m$-th atomic 
state is given by $\cE_m\simeq K_z^2/(2M)-0.16\,l_m^2
+E_{\perp m}(K_\perp)+mb/M+(1+\sigma_z)b/(2M)$, where the proton 
spin $\sigma_z=\pm 1$ (compare with Eq.~[3.27]).}
\be
z_m(H)\simeq \left(1+e^{-b/MT}\right)
\sum_{m=0}^{\infty}{M_{\perp m}'\over M_\perp'}
\exp\left[-{1\over T}\left(0.16\,l^2-0.16\,l_m^2+m{b\over M}\right)\right],
\ee
where 
\be
M_{\perp m}'=M_{\perp m}\left(1-{2M_{\perp m}T\over \tau_m}\right)^{-1},
\ee
so that $M_{\perp 0}=M_{\perp}$ (cf.~Eq.~[5.4]), and 
$M_{\perp m}$, $\tau_m$ are given in Eqs.~(3.12) and (3.28). 
The sum in Eq.~(5.7) simplifies in two extreme regimes of the field strength:
For $b<<b_{crit}$ one may need to include a number of terms in the sum, but 
then $M_{\perp m}'$ is close to $M$; For $b>>b_{crit}$ the effective mass
$M_\perp'$ is much larger than $M$, but since $b>>MT$ one need to include 
only the $m=0$ ground state, and hence $z_m(H)=1$. 

Finally, it is instructive to consider the partition function based on 
the alternative scheme discussed in Sec.~IV. Using Eq.~(4.9)
(and relacing $b/2.3$ by $b$; see the discussion following
Eq.~[4.10]), we see that 
$Z(H)$ can still be written as Eq.~(5.1), but the
transverse partition function $Z_\perp$ is now given by
\be
Z_{\perp} \sim {\cA_gb\over 2\pi}\sum_{m_t=0}^{m_{t,max}}
\exp\left[-{l\over T}\ln (1+C'm_t)\right]
\sim {\cA_gb\over 2\pi}\int_0^{m_{t,max}}dm_t (1+C'm_t)^{-l/T},
\ee
where the factor $A_gb/2\pi$ comes from the Landau degeneracy of 
the $m_t$-th state, and $m_{t,max}\sim R_g^2b$. Clearly, 
Eq.~(5.9) has the same form as Eq.~(5.3)
in the limit of $b/M>>l$, demonstrating the equivalence of the two
energy level schemes of Sec.~II-III and Sec.~IV. 

\subsection{\bf Saha Equation}

The partition function of the free electrons in the ground Landau level
is given by 
\be
Z(e)\simeq R_g\left({T\over 2\pi}\right)^{1/2}
\left(\cA_g{b\over 2\pi}\right),
\ee
where the factor $R_g(T/2\pi)^{1/2}$ represents the free  $z$-motion and 
the factor $\cA_g b/(2\pi)$ is the Landau degeneracy. 
For the free protons (see footnote 1), we have
\begin{eqnarray}
Z(p) &=& R_g\left({MT\over 2\pi}\right)^{1/2}
\left(\cA_g {b\over 2\pi}\right)
\sum_{n_2=0}^{\infty}g_{n_2}\exp\left(-{n_2b\over MT}\right)\nonumber\\
&=& V_g\left({MT\over 2\pi}\right)^{1/2}
{b\over 2\pi}\left[\tanh\left({b\over 2MT}\right)\right]^{-1},
\end{eqnarray}
where the sum extends over all 
Landau levels of the proton, and $g_{n_2}$ is the spin degeneracy: 
$g_{n_2}=1$ for $n_2=0$ and $g_{n_2}=2$ for $n_2>0$. 
Given $Z(e)$, $Z(p)$ and $Z(H)$, the ionization-recombination
equilibrium can be obtained using the condition
$\mu(e)+\mu(p)=\mu(H)$ for the chemical potentials.
In the density and temperature regimes of interest, 
with $T<<K_{\perp p}^2/(2M_\perp)$, we have
\be
{X(H)\over X_pX_e}\simeq n_g\left({b\over 2\pi}\right)^{-2}\,
M_\perp\left(1-{2M_\perp T\over \tau}\right)^{-1}
\left({T\over 2\pi}\right)^{1/2}\tanh\left({b\over 2MT}\right)
\exp\left({|E(H)|\over T}\right)z_m(H),
\ee
where $X(H)=n(H)/n_g$, $X_p=n_p/n_g$, $X_e=n_e/n_g$ are the number 
density fraction of different species, 
$M_\perp=M+\xi b/l$ (with $\xi\simeq 2.8$), 
and $z_m(H)$ is given by Eq.~(5.7). 
This is the generalized Saha equation in the presence of a superstrong 
magnetic field. More details on the applications of this result to 
neutron star atmospheres will be presented elsewhere\cite{Lai95b}.

\section{\bf SUMMARY}

The effects of center-of-mass motion of neutral hydrogen 
atom in a strong magnetic field are rather intricate, mainly 
due to the high degree of degeneracy associated with the quantum states. 
Using the usual pseudomomentum scheme (Sec.~II-III),
we have obtained approximate solutions for the energy of the atom
as a function of the field strength and conserved pseudomomentum
for a wide range of parameter regimes. In particular, we have 
considered field strengths $B\go B_{crit}\sim 4\times 10^{13}$ G, 
when the Landau excitation energy of proton is considerable. 
States with large transverse pseudomomentum have small binding 
energy and transverse velocity, but are nevertheless
quite distinct from fully ionized states.
We have concentrated on convenient analytic fitting formulae which
give at least a reasonable approximation over various parameter
regimes (see particularly Eq.~(3.26) for the ``transverse kinetic
energy''). Since there may be neutron star atmospheres with 
$B\sim 10^{13}-10^{14}$ G, we are particularly interested in 
the cases with $B>>B_{crit}$, where the proton Landau energy 
$\hbar\omega_p$ is very large. By considering an alternative scheme 
to the usual pseudomomentum formulation (Sec.~IV), we have shown 
that there are atomic states with orbital wavefunctions 
orthogonal to that of the ground state, but without any Landau
excitation appearing in their energies.

We have also derived the generalized Saha equation for the equilibrium
between neutral hydrogen atoms and the ionized component. 
We focused on the cases of astrophysical interest, where
the density is relatively low and the thermal energy $k_BT$ is 
small compared to the atom's ground state binding energy. 
Although the maximum transverse velocity of bound atoms
is small in strong magnetic fields, the statistical weight due to 
transverse motion is actually increased by the strong fields, not 
decreased (Sec.~V). The statistical weight of highly excited bound
states is smaller than that of the fully ionized component.
Our results are important for determining the 
physical conditions of magnetic neutron star atmospheres as well as 
the soft X-ray (or EUV) radiation spectra from them. Some of these issues
will be studied in a future paper\cite{Lai95b}.

\acknowledgments
We thank Lars Bildsten, George Pavlov and Ira Wasserman for useful 
discussions. This work has been supported in part by NSF Grants 
AST 91--19475, AST 93--15375, and NASA Grant NAGW--666 
to Cornell University, and NAGW-2394 to Caltech. 

\appendix
\section{}

In this Appendix, we derive some mathematical formulae
needed for evaluating the function $G_{m_1m_1'}^{(m_t)}(z)$
defined by Eq.~(4.8).  
(Here the lengthscale is in units of the cyclotron radius 
$\hat\rho$.) Using the identity 
\begin{equation}
{1\over r}={1\over 2\pi^2}\int\! {d^3q\over q^2}e^{i{\bf q}\cdot\br},
\end{equation}
we write Eq.~(4.8) as 
\be
G_{m_1m_1'}^{(m_t)}(z)=G_{m_1'm_1}^{(m_t)}(z)=
{1\over 2\pi^2}\int\!{d^3q\over q^2}e^{iq_zz}
\langle m_t+m_1|e^{i{\bf q}_\perp\cdot\br_{1\perp}}|m_t+m_1'\rangle
\langle m_1'|e^{-i{\bf q}_\perp\cdot\br_{2\perp}}|m_1\rangle.
\ee
Using the general result\cite{Virtamo75} for the  matrix element
$\langle m'|e^{i{\bf q}_\perp\cdot\br_\perp}|m\rangle$, and 
integrating over $dq_z$, we obtian
(assuming $m_1\ge m_1'$ without loss of generality): 
\be
G_{m_1m_1'}^{(m_t)}(z)=\int_0^\infty\! dq\, e^{-q|z|-q^2}
\left({q^2\over 2}\right)^{m_1-m_1'}\!\!
\sqrt{(m_t+m_1')!\,m_1'!\over (m_t+m_1)!\,m_1!}\,
L_{m_t+m_1'}^{m_1-m_1'}\!\!\left({q^2\over 2}\right)
L_{m_1'}^{m_1-m_1'}\!\!\left({q^2\over 2}\right),
\ee
where $L_n^m$ is the Laguerre polynomial of
order $n$\cite{Abramowitz72}.
We now define constant coefficients $g_n^{(m_t)}(m_1,m_1')$ via
\be
\left({x\over 2}\right)^{m_1-m_1'}\!\!
\sqrt{(m_t+m_1')!\,m_1'!\over (m_t+m_1)!\,m_1!}\,
L_{m_t+m_1'}^{m_1-m_1'}\!\!\left({x\over 2}\right)
L_{m_1'}^{m_1-m_1'}\!\!\left({x\over 2}\right)
=\sum_{n=0}^{m_1+m_1'+m_t}\! g_n^{(m_t)}(m_1,m_1')L_n(x).
\ee
Using the relation
\be
V_n(z)=\int_0^\infty\!\!dq
e^{-q^2/2-q|z|}L_n\biggl({q^2\over 2}\biggr)
={1\over\sqrt{2}\, n!}\int_0^{\infty}\!\! dx
{x^n e^{-x}\over (x+z^2/2)^{1/2}},
\ee
where the second equality can be used to evaluate the function
$V_n(z)$, we obtain
\be
G_{m_1m_1'}^{(m_t)}(z)=\sum_{n=0}^{m_1+m_1'+m_t}\! 
{1\over\sqrt{2}}\,g_n^{(m_t)}(m_1,m_1')
\,V_n\!\left({z\over\sqrt{2}}\right).
\ee
We calculate the coefficients $g_n^{(m_t)}(m_1,m_1')$ using the
orthogonal relations of Laguerre polynomials. We can identify
two special cases: (i) When $m_t=0$, we have 
$G_{m_1m_1'}^{(0)}(z)=E_{m_1m_1'}(z)$ and 
$g_n^{(0)}(m_1,m_1')=e_n(m_1,m_1')$, where 
$E_{m_1m_1'}(z)$ and $e_n(m_1,m_1')$ are given in\cite{Lai92};
(ii) When $m_1=m_1'=0$, we have 
$G_{00}^{(m_t)}(z)=D_{0m_t}(z)$ and 
$g_n^{(m_t)}(0,0)=d_n(0,m_t)$, where 
$D_{m_1m_1'}(z)$ and $d_n(m_1,m_1')$ are again given 
in\cite{Lai92}. 



\begin{figure}
\caption
{Numerical results for the energies of
hydrogen atom in the $m=0$ state (solid lines) and $m=1$ state 
(dotted lines) as a function of $R_K$, obtained by solving 
Eq.~(3.19). The upper curves are for $B_{12}=10$, and the 
lower curves for $B_{12}=100$.
}
\end{figure}

\begin{figure}
\caption
{The energy eigenvalue $\cE^{(m_t)}$ calculated from Eq.~(4.7)
for the $m_t=0$ (upper panel) and $m_t=1$ (lower panel) states.
The ratio $\cE^{(m_t)}/E_o$, where $E_o=-0.16\,A\, l^2$ is 
the ground-state energy as given by Eq.~(3.2), is plotted
against $m_{1 max}$, the maximum values of $m_1$ in the
sum of Eq.~(4.6). The filled triangles are for $B_{12}=100$,
the open circles for $B_{12}=1000$, and the filled circles for
$B_{12}=5000$. 
}
\end{figure}


\begin{references}

\bibitem{Ruderman74}
M. Ruderman, in {\it Physics of Dense Matter} (I.A.U. Symp. No. 53),
Edited by C.~J. Hansen (Dordrecht: Holland, 1974).

\bibitem{Ruder94}
H. Ruder et al. {\it Atoms in Strong Magnetic Fields}
(Springer-Verlag, 1994).

\bibitem{Elliot60}
Early study of essentially the same problem dealt with 
excitons in semiconductors,
where the strong field condition can be mimicked in virtue of the 
small effective mass and the presence of a dielectric medium. See
R.~J. Elliot and R. Loudon, J. Phys. Chem. Solids 
{\bf 15}, 196 (1960); H. Hasegawa and R.~E. Howard, {\em ibid.}
{\bf 21}, 179 (1961).

\bibitem{Cohen70}
For calculations in the superstrong field regime ($B>>10^{9}$ G), see 
R. Cohen, J. Lodenquai, and M. Ruderman, Phys. Rev. Lett. {\bf 25}, 
467 (1970); V. Canuto and D. C. Kelley, Astrophys. Space Sci.
{\bf 17}, 277 (1972); J. Simola and J. Virtamo, J. Phys. B: At. Mol. Phys. 
{\bf 11}, 3309 (1978), and references therein.

\bibitem{Garstang77}
For a review of the intermediate field regime ($B\lo 10^9$ G), see
R.~H. Garstang, Rep. Prog. Phys. {\bf 40}, 105 (1977). 

\bibitem{Rosner84}
Hydrogen atoms in arbitrary magnetic fields were first studied by
W. R\"osner, G. Wunner, H. Herold and H. Ruder,
J. Phys. B: At. Mol. Phys. {\bf 17}, 29 (1984).

\bibitem{Goldman91}
For high-precision calculations of the atomic binding energy, including 
the relativistic effects, see 
S.~P. Goldman and Z. Chen, Phys. Rev. Lett. {\bf 67}, 1403 (1991),
and references therein.

\bibitem{Lai92}
D. Lai, E.~E.~Salpeter and S.~L.~Shapiro, 
Phys. Rev. A {\bf 45}, 4832 (1992).

\bibitem{Vincke92}
M. Vincke, M. Le Dourneuf and D. Baye, J. Phys. B: At. Mol. Phys. 
{\bf 25}, 2787 (1992).

\bibitem{Potekhin94}
A.~Y. Potekhin, J. Phys. B: At. Mol. Phys. {\bf 27}, 1073 (1994).

\bibitem{Lai95}
D. Lai and E.~E.~Salpeter, Phys. Rev. A, in press (1995). 

\bibitem{Khersonskii87}
V. K. Khersonskii, Sov. Astron. {\bf 31}, 225 (1987).

\bibitem{Miller92}
M.~C. Miller, Mon. Not. Roy. Astron. Soc. {\bf 255}, 129 (1992).

\bibitem{Ventura92}
J. Ventura, H. Herold, H. Ruder and F. Geyer, Astron. \& Astrophys., 
{\bf 261}, 235 (1992).

\bibitem{Pavlov94}
G.~G. Pavlov, Y.~A. Shibanov, V.~E. Zavlin and R.~D. Meyer, 
in {\it Lives of Neutron Stars}, ed.~A. Alpar, U. Kiziloglu and 
J. van Paradijs (Kluwer, Dordrecht, 1994).

\bibitem{Lamb52}
W.~E. Lamb, Jr., Phys. Rev. {\bf 85}, 259 (1952).

\bibitem{Gorkov68}
L.~P. Gor'kov and I.~E. Dzyaloshinskii, Sov. Phys. JETP {\bf 26}, 449 
(1968).

\bibitem{Avron78}
J.~E. Avron, I.~B. Herbst and B. Simon, Ann. Phys. (NY) {\bf 114}, 
431 (1978).

\bibitem{Herold81}
H. Herold, H. Ruder and G. Wunner, J. Phys. B: At. Mol. Phys. 
{\bf 14}, 751 (1981).

\bibitem{Johnson83}
B.~R. Johnson, J.~O. Hirschfelder, and K.-H. Yang, 
Rev. Mod. Phys. {\bf 55}, 109 (1983). 

\bibitem{Schmel88}
Born-Oppenheimer approximation in strong
magnetic field is discussed in 
P. Schmelcher, L. S. Cederbaum and H.-D. Meyer,
Phys. Rev. A., {\bf 38}, 6066 (1988).

\bibitem{Vincke88}
M. Vincke and D. Baye, J. Phys. B: At. Mol. Phys. {\bf 21}, 2407 (1988);
D. Baye and M. Vincke, {\it ibid}, {\bf 23}, 2467 (1990).

\bibitem{Pavlov93}
G. G. Pavlov and P. Meszaros, Astrophys. J. {\bf 416}, 752 (1993).

\bibitem{Pavlov94b}
G. G. Pavlov et al. Astron. \& Astrophys. {\bf 289}, 837 (1994).

\bibitem{Ventura94}
J. Ventura, H. Herold, \& Kopidakis,
in {\it Lives of Neutron Stars}, ed.~A. Alpar, U. Kiziloglu and 
J. van Paradijs (Kluwer, Dordrecht, 1994).

\bibitem{Shabad86}
A.~E. Shabad and V.~V. Usov, Astrophys. Space Sci.
{\bf 128}, 377 (1986). 

\bibitem{Paczynski92}
B. Paczy\'nski, Acta Astron., {\bf 42}, 145 (1992). 

\bibitem{Duncan93}
R.~C. Duncan and C. Thompson, Astrophys. J., {\bf 392}, L9 (1993). 

\bibitem{Chakrabarty94}
D. Chakrabarty, et al., in {\it Proc.~of the Second Compton
Symposium}, Edited by C.~E. Fichtel, N. Gehrels and J.~P. Norris
(AIP Press: New York, 1994).


\bibitem{Neuhauser87}
D. Neuhauser, S.~E. Koonin, and K. Langanke, Phys. Rev. A
{\bf 36}, 4163 (1987).

\bibitem{Angelie78}
C. Angelie and C. Deutch, Phys. Lett. {\bf 67A}, 353 (1978).

\bibitem{Landau77}
L.~D. Landau and E.~M. Lifshitz, {\it Quantum Mechanics}
(Oxford: Pergamon Press, 1977).

\bibitem{Haines69}
L.~K. Haines and D.~H. Roberts, Am. J. Phys. {\bf 37}, 1145 (1969). 

\bibitem{Abramowitz72}
M. Abramowitz and I. Stegun, {\it Handbook of Mathematical 
Functions} (New York: Dover) (1972).


\bibitem{Press87}
W.~H. Press, B.~P. Flannery, S.~A. Teukolsky and W.~T. Vetterling,
{\it Numerical Recipes: The Art of Scientific
Computing} (Cambridge Univ. Press, 1987).

\bibitem{Lai95b}
D. Lai and E.~E.~Salpeter, Astrophys. J., in preparation (1995).

\bibitem{Virtamo75}
J. Virtamo and P. Jauho, Il Nuovo Cimento, {\bf 26B}, 537 (1975);
A. A. Sokolov and I. M. Ternov, {\it Synchrotron Radiation}
(Oxford: Pergamon Press) (1968).

\end{references}
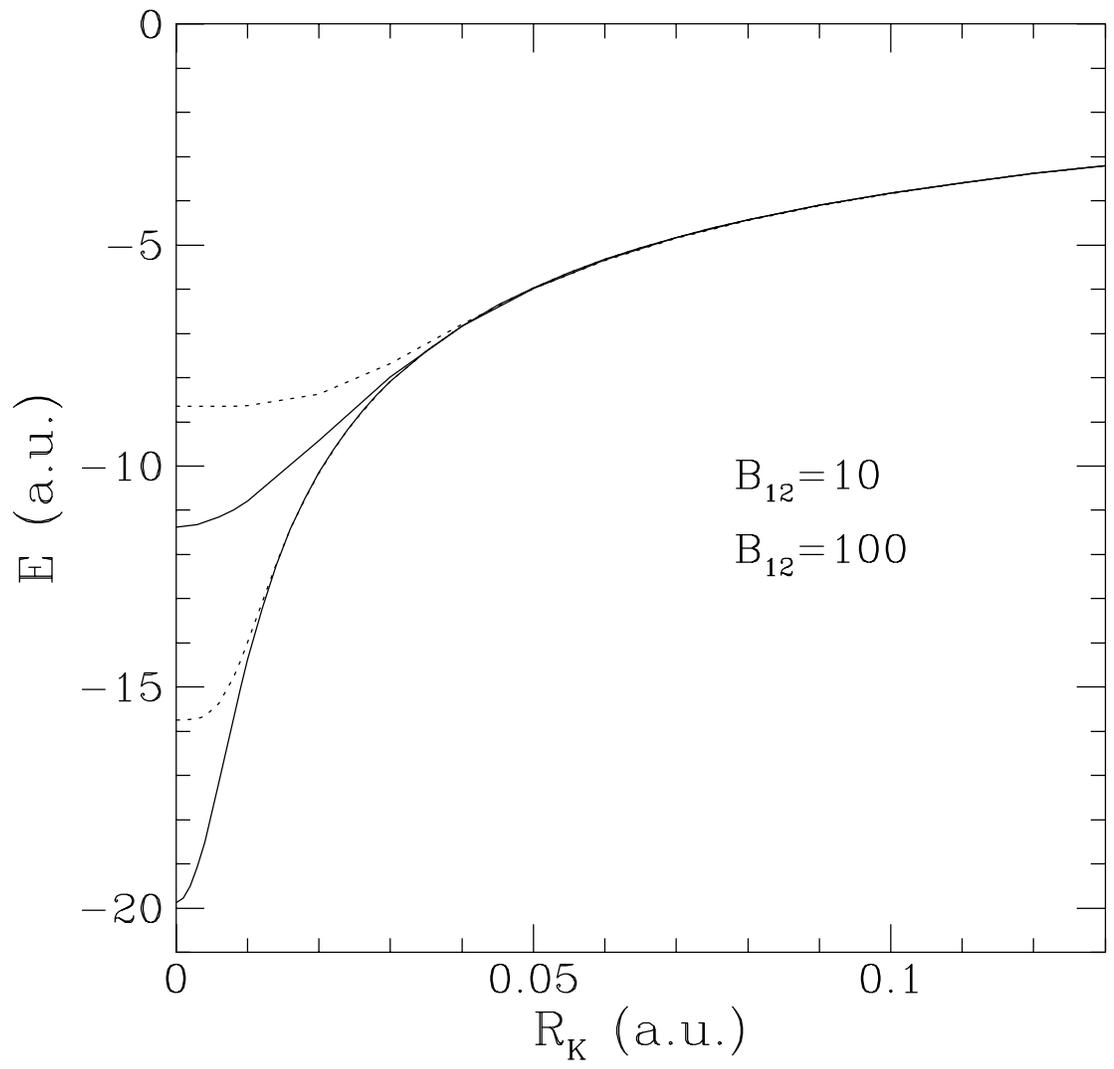
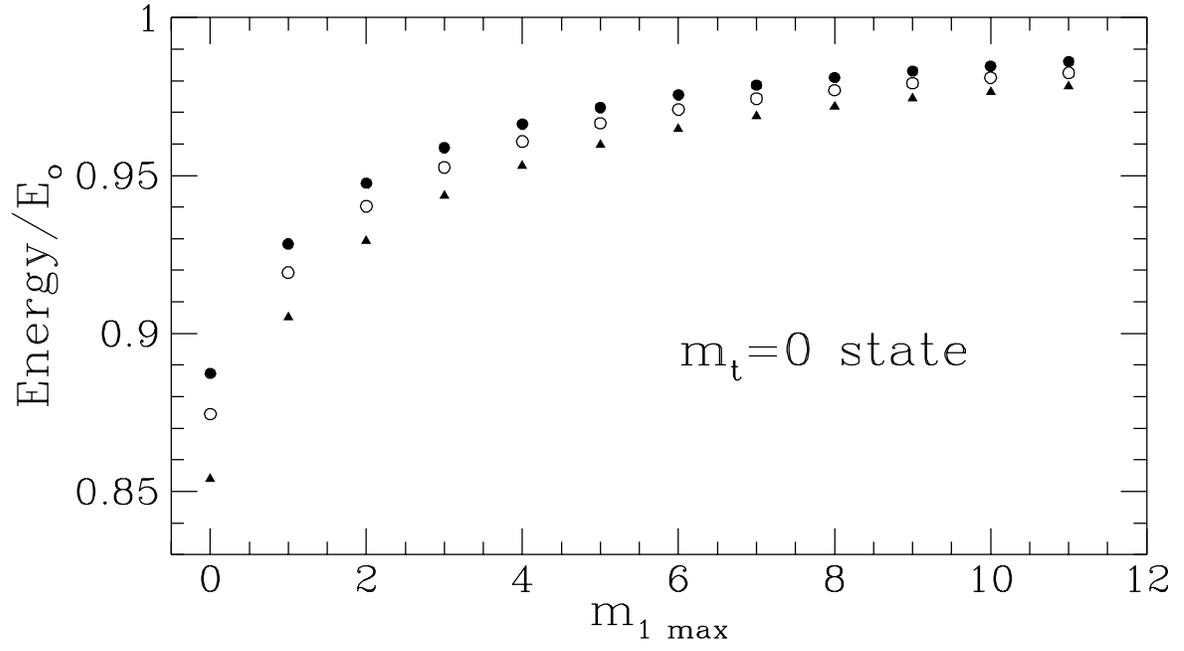
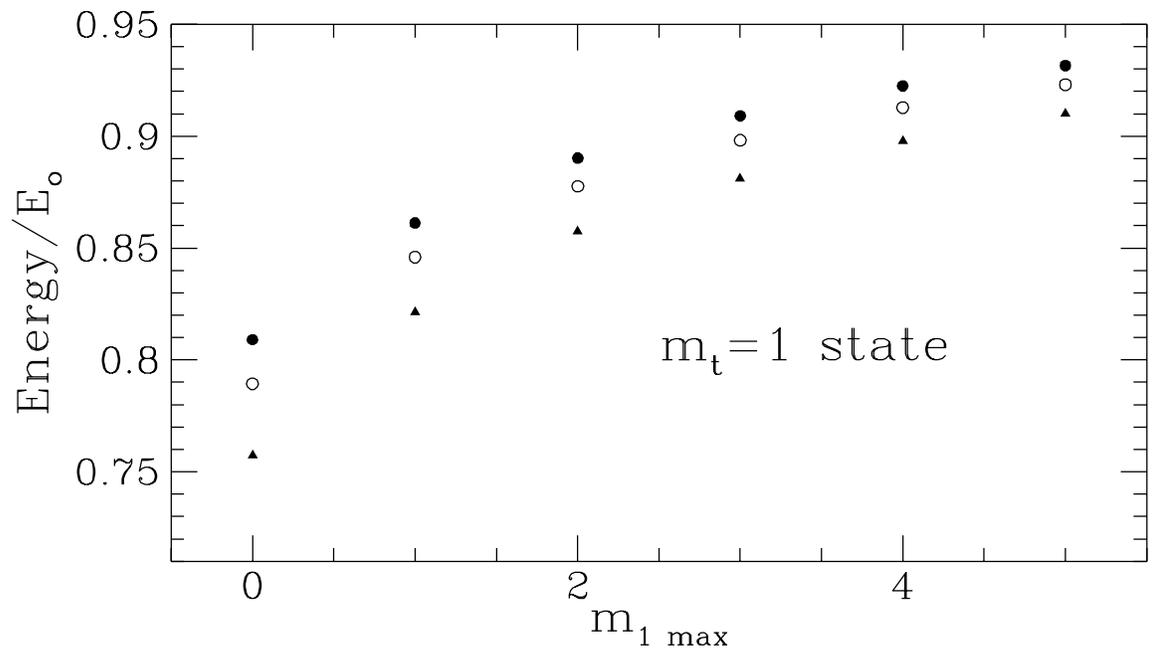
\end{document}